\documentstyle[prl,tighten,aps]{revtex}
\def\beqa{\begin{eqnarray}}
\def\eeqa{\end{eqnarray}}
\def\beq{\begin{equation}}
\def\eeq{\end{equation}}

\def\umunu{^{\mu\nu}}
\def\dmunu{_{\mu\nu}}

\def\bib#1{$^{\ref{#1}}$}

\def\ie{{\it i.e. }}
\def\eg{{\it e.g. }}

\def\a{\`a }

\def\l{\cal L}
\begin{document}
\def\bib#1{[{\ref{#1}}]}

	 \title{Quantum Signature of Cosmological Large Scale 
                Structures}

\author{Salvatore Capozziello$^{a,c,}$\thanks{
E-mail:capozziello@vaxsa.csied.unisa.it}, 
Salvatore De Martino$^{b,c,}$\thanks{E-mail:demartino@physics.unisa.it},
Silvio De Siena$^{b,c,d}$\thanks{E-mail:desiena@physics.unisa.it} 
and Fabrizio Illuminati$^{b,c,d}$\thanks{E-mail:
illuminati@vaxsa.csied.unisa.it} \\ 
{\em $^a$Dipartimento di Scienze Fisiche ``E. R. Caianiello",}\\ 
{\em $^b$Dipartimento di Fisica,}\\ 
{\em $^c$INFN, Sez. di Napoli and INFM Unit\a di Salerno, \\
Universit\`a di Salerno, I-84081 Baronissi (SA) Italy,} \\
{\em $^d$Fakult\"{a}t f\"{u}r Physik, Universit\"{a}t Potsdam, \\
     Am Neuen Palais 19, D--14469, Potsdam, Germany.}}
\date{1 September 1998}
\maketitle

	      \begin{abstract}
We demonstrate that to all large scale
cosmological structures where gravitation is 
the only overall relevant interaction 
assembling the system (\eg galaxies), 
there is associated a characteristic
unit of action per particle whose order of magnitude 
coincides with the Planck action constant $h$.
This result extends the class of physical systems
for which quantum coherence can act on
macroscopic scales (as \eg in superconductivity) 
and agrees with the absence of screening mechanisms for 
the gravitational forces, as predicted by some 
renormalizable quantum field theories of gravity. 
It also seems to support those lines of thought invoking that
large scale structures in the Universe should be connected to
quantum primordial perturbations as requested by inflation, 
that the Newton constant should vary with time and distance and, 
finally, that gravity should be considered as an effective
interaction induced by quantization. 

               \end{abstract}

\vspace{20. mm}

\vspace{20. mm}

\section{\normalsize \bf Introduction}

Explaining the large scale structure of the Universe is one of the hardest 
task of modern cosmology since the growing amount of observations seems to 
escape any coherent scheme able to connect all the parts of the puzzle.

Essentially, from the fundamental physics point of view, we would like to 
reconduct cosmic structures and their evolution to some unifying theory in 
which all the today observed interactions are treated under the same 
standard. In this case, what we observe nowaday
on macroscopic, astrophysical scales would be just a consequence
of quantum fluctuations at early epochs. Then, we should seek for some
``enlarging'' mechanism which after one (or more than one) symmetry
breaking would be capable of yielding 
structures like galaxies from primordial
quantum spectra of perturbations. 

The so called ``inflationary paradigm'' \bib{kolb} related to the several
unifying theories (\eg superstrings, GUT, SUSY, and so on) should be
succesful if some ``experimentum crucis'' would select the right 
model.

On the other hand, particle physicists need cosmological  predictions
and observations since the energies for testing unified theories
are so high that it is extremely unlikely they will be ever 
reached on earth--based laboratories. 

As a matter of fact, cosmology needs particle physics and vice versa.
The point is that remnants of primordial epochs should be found by
cosmological observations and, by them, one should constraint 
elementary particle physics models. 

This philosophy has been pursued by several researchers; 
first of all by Sakharov \bib{sakharov} who in 1965 argued 
that quantum primordial
fluctuations should have expanded towards the present epoch 
leading first to classical energy--density perturbations and, 
after the decoupling from
the cosmological background, to the observed galaxies, clusters and
superclusters of galaxies. Shortly, the underlying issue 
of any modern theory
of cosmological perturbation is this: 
primordial quantum fluctuations should
be enlarged by cosmological dynamics to the present large scale structures.
Now the problem is not only whether observations agree with this scheme
(\eg COBE and IRAS data or large scale structure surveys \bib{hucra}) but,
mainly, whether the astrophysical and cosmological systems ``remember''
their quantum origin or not. 

In some sense, the question becomes philosophical: 
It is well known that quantum and classical domains are distinct; 
the frontier is marked by
the Planck constant $h$
which separates microscopic from macroscopic scales. If the lengths, the
energies, the times become much larger than $h$, 
we are in the classical
physics regime with likely little hope of recovering 
quantum signatures.

However, despite of this apparent sharp division of the classical and
quantum worlds, macroscopic quantum
phenomena exist and some behaviors of
classical systems can be explained only in the 
framework of quantum mechanics. 
The high $T_{c}$ superconductivity and several other
macroscopic coherent systems (\eg optical fibres) 
are famous instances of these peculiar phenomena
in which a quantum ``memory'' persists at the macroscopic
scale.

Recently, a new intriguing conjecture has been
proposed to find signatures of $h$ at the classical, 
macroscopic scales: Francesco Calogero has argued about 
a possible gravitational origin of quantization
emerging thanks to the universal interaction 
of every particle in the Universe with the gravitational
stochastic background field 
generated by all other particles \bib{calogero}. 

In the framework of this fluctuative Machian scheme, 
Calogero has been able to show how classical nonlinearity
and chaoticity of the gravitational interaction in the Universe
yields a unit of action per particle that coincides, in order
of magnitude, with $h$.

Further studies \bib{demartino} have generalized 
the scheme of Calogero to the other fundamental interactions
responsible for large scale macroscopic structures,
finding that $h$ is the characteristic action per particle
also for macroscopic systems not bound by gravitational interactions
but by other forces, \eg electromagnetic.
 
In this new scheme, classical laws of force $F(R)$ describing the 
interactions among the constituents of $N$--particle systems of 
mean length scale $R$, lead to $h$ as the characteristic action
per particle. 
The forces $F(R)$ considered can be, for instance, 
the electromagnetic interactions between charged particles
in large macroscopic systems as charged beams in particle
accelerators, plasmas, and neutral dipolar crystals, or 
the strong interactions between quarks in hadronic bound aggregates, 
and so on. In other words, $F(R)$ needs not necessarily have to be 
the gravitational interaction.

The conclusion seems to be that the space--time scales
of several mesoscopic and macroscopic coherent
aggregates are ruled by characteristic actions of order $h$.
From this point of view, gravitation loses the privileged status
of ``origin'' of quantization that it played in the original 
scheme of Calogero \bib{calogero}, while the central result of
his investigation, as generalized by De Martino {\it et al.}
\bib{demartino}
is that it provides a method to find some 
quantum mechanical signature
in classical macroscopic structures, and to connect it
to the observed measured values of their space and time
scales.

Having such a procedure at hand, it seems then very
natural and appealing to investigate whether it can
be applied to determine the existence of unambiguous
quantum mechanical signatures or ``memories'' for
large scale cosmological structures. This is exactly
the issue considered in the present paper: in particular,
to see whether it is possible to
explain the  large scale cosmological structures 
by finding $h$ as the characteristic unit of action for systems 
as galaxies, clusters of galaxies, and super clusters of galaxies.

As we shall see below, several renormalizable 
quantum theories of gravity require a 
modification, in the low energy limit, of Newton law. 
Furthermore, if we do not require enormous amounts of dark matter 
as the only mechanism to explain the puzzle of the present day
astrophysical observations, a scale--dependent gravitational 
interaction is also needed. 

In this framework, we will show that the method by Calogero
in the generalized formulation of De Martino {\it et al.}
\bib{demartino}
can be successfully applied to 
macroscopic systems where the overall relevant interaction is 
gravitation (that is where the size of the bound 
systems is determined by gravity alone) and that in all cases
the characteristic unit of action per constituent is
$h$. Therefore the original result of Calogero holding
for the entire Universe is recovered also for cosmological
structures at smaller scales, and their 
quantum signatures become evident.

The paper is organized as follows.
In Sec. 2, we briefly review why varying effective 
gravitational couplings
and non--Newtonian effective potentials can avoid several 
shortcomings in fundamental elementary physics and cosmology.
Sec. 3 is devoted to the discussion of the method of Calogero
generalized by De Martino {\it et al.} to determine the characteristic
minimal unit of action per constituent in classical 
macroscopic systems. 
In Sec. 4, we determine the characteristic minimal unit of 
action per constituent in the case of
gravitational large scale cosmological
structures and find, also in this case, that it is $h$.
A major role is played by the spatial scale of the structure, 
by the ratio of its mass with the number of baryons present in it,
and by the space--time variation of $G_{N}$.
Conclusions are drawn in Sec.5. 

\section{\normalsize \bf The variation of $G_{N}$ and the non--Newtonian 
effective potentials}

The possibility of considering a variable Newtonian coupling constant 
is, at least, sixty years old. 
In 1937 Dirac \bib{dirac} put forward his so called 
Large Numbers Hypothesis \bib{barrow} 
in which some intriguing numerical coincidences such as that of the 
ratio of the electromagnetic to gravitational force with 
the number of protons in the Universe and with 
the age  of the Universe could be explained in the framework  
of some unified theory of the micro-- and macrophysics. 
In order to keep the constant values of
$e$, the electron charge, $m_{e}$, the electron mass, and $m_{p}$, the
proton mass, Dirac asked for a variation\footnote{From now on,
$G_{N}=6.67\times 10^{-8}$g$^{-1}$cm$^{3}$s$^{-2}$ 
denotes the Newton constant
measured by Cavendish--like experiments;
 $G(r,t)$ denotes
a variable gravitational coupling whose 
possible explicit forms will be 
given below. In general, such a coupling can depend also on 
a space--time dependent scalar
field $\phi(r,t)$ \bib{cimento}.} of $G$ of the form
\beq
\label{1}
G\sim\frac{1}{t}\,.
\eeq
In this hypothesis, the gravitational strength 
goes to zero for large times.
Similar arguments apply also to the cosmological constant and, 
in the framework 
of the Large Numbers Hypothesis, it is argued by several
authors (see, for instance, \bib{berman}, \bib{beesham})
that 
\beq
\label{2}
\Lambda\sim\frac{1}{t^2} \, .
\eeq

A time--dependent gravitational coupling was conceived also by Sciama
\bib{sciama} and Jordan \bib{jordan} who provided further arguments  
supporting this view. 
In the Brans--Dicke approach \bib{brans},  
General Relativity was modified by introducing a scalar field in the 
equations of motionto to make them consistent with Mach principle. 
Such a consistency is in fact achieved if the
gravitational coupling is a variable quantity.

More recently, the so--called induced gravity theories 
\bib{smolin} have inquired the possibility  
that the gravitational and cosmological constants 
may not be phenomenological parameters to be introduced
by hand, but they might be rather induced
from the spontaneous symmetry breaking of a scalar
field $\phi$ nonminimally coupled to the Ricci scalar ${\cal R}$ 
in the interaction Lagrangian \bib{cqg}. 
The resulting gravitational effective 
action will contain higher--order terms in 
the geometrical invariants like
${\cal R}^2$ or ${\cal R}\umunu {\cal R}\dmunu$ 
\bib{stelle},\bib{fradkin},\bib{avramidy},
nonminimally coupled terms like
$\xi\phi^2 {\cal R}$ \bib{smolin},\bib{birrell},
or nonminimally--coupled--higher--order 
terms like $\xi\phi^2 {\cal R}^2$
\bib{amendola}. 
Such theories are renormalizable at 
one--loop level when graviton--graviton or matter--graviton interactions
are considered. To achieve such a result, two effective 
running constants must be
renormalized, $G_{eff}$ and $\Lambda_{eff}$, 
which, in the low energy limit,
reduce to $G_{N}$ and $\Lambda$. 

From a cosmological point of view, this kind of theories can give rise to
inflationary behaviours solving the shortcomings of 
the standard cosmological
model \bib{la},\bib{capozziello}. 
The mechanism giving rise to the induced gravitational
interaction resembles that of 
vacuum polarization in QED, but with some relevant 
differences. 
The vacuum energy of such theories is very small but the 
net effect of quantum fluctuations can provide 
sizeable corrections at all
scales of distances since 
gravity coherently couples with itself and with any
form of matter. 
Besides, the absence of screening mechanism for gravitons
\bib{weinberg} allows that quantum gravity effects can play a role 
also at macroscopic and cosmological scales \bib{goldman}. 
Although these one--loop
quantum gravity theories are known to exhibit pathological
behaviors with respect to unitarity (ghost poles in the 
tree--level propagators), 
their breakdown is expected near the Planck
scale (epoch)
while larger scales are nonsensitive to this shortcoming. 
Thus, these
theories can be considered the effective theories of gravity valid 
at length
scales much larger than the Planck length.

One of the main results of this approach is that the 
renormalization group
equation for the gravitational coupling can be analyzed
in some detail  
\bib{avramidy},\bib{gellmann},\bib{tonin},
and, depending on the values of the parameters 
and on the momentum scales considered 
in studying the behavior of the $\beta$
functions, $G_{eff}$ increases or 
decreases with the distance.
Then, in general, we can write
\beq
\label{3}
G(r)=G_{N}f(r)\,,
\eeq
where $G_{N}$, as we have said, is the Newton constant as measured in 
laboratory. 

If we are dealing with higher--order theories of the form 
\bib{avramidy}
\beq
\label{5}
{\cal A}=\int d^4x\sqrt{-g}\left\{\Lambda-\frac{{\cal R}}{16\pi G_{N}}
+c_{1}{\cal R}^2
+c_{2}{\cal R}\dmunu {\cal R}\umunu+\cdots\right\}\,,
\eeq
one can show that a consistent choice is
\beq
\label{4}
G(r)=\chi G_{N}\left(\frac{r}{r_{0}}\right)^{\eta}
\ln\left(\frac{r}{r_{0}}\right)\,,
\eeq
where the parameters $\eta$, $\chi$, and $r_{0}$ 
depend on the details of the
model \bib{bertolami}.  
 
Another approach \bib{fradkin} exploits a 
technique which treats the higher
derivative terms as if they incorporated 
additional massive bosons, yielding 
Yukawa--like effects in the Newtonian potential in the
low energy limit. In this case one has
\beq
\label{6}
V(r)=-\frac{G(r)M}{r}\,,
\eeq
where $G(r)$ can take the form
\beq
\label{7}
G(r)=G_{N}\left(1+a_{0}e^{-r/{r_{0}}}\right)\,,
\eeq
or the form
\beq
\label{8}
G(r)=G_{N}\left(1+d_{1}e^{-m{_{1}}(r)r}+ d_{2}e^{-m{_{2}}(r)r}\right)\,,
\eeq
as shown by Stelle \bib{stelle}. 

Some authors \bib{sanders},\bib{eckhardt}
have exploited potentials like (\ref{7}) 
in order to explain the flat 
rotation curves of spiral galaxies. 
Their approach is phenomenological and
the adjustments of parameters is often ``ad hoc'' in order to fit
the experimental data.
  
Yukava--like corrections result also if one deals with galaxies as potential
wells \bib{3kpc}. In this case, the theory of Newtonian perturbations on a
background fluid is sufficient to explain, in a quite simple way, the 
dynamics of hot components of galaxies like the bulge.
In all these approaches, the amount of dark matter still
required to match observations is greatly suppressed compared 
to the amount needed to ``cure'' the standard cosmological
model.

From the above discussion, it appears not
so unnatural to allow
for a running gravitational coupling. 
Such a variation can be with 
respect to the time (\eg in cosmology) or, 
in general, with respect to the scale. 

In the present paper we show that considering a 
scale--dependent $G(r)$, it is 
possible to determine, exploiting the
results of Calogero \bib{calogero}
and of De Martino {\it et al.} \bib{demartino},
a characteristic minimal action 
of the order of Planck
constant $h$ for any system bound only by gravity.
This is the case for galaxies, clusters, and superclusters of galaxies
(On the other hand, we exclude stars from our analysis, since in this
case electromagnetic and nuclear interactions play 
a crucial role in the binding of the system).

\section{\normalsize \bf The characteristic unit of 
action for macroscopic systems}

As we have seen in the previous section, 
the issue of a running gravitational 
coupling is relevant
in the context of quantum theories of gravity and
quantum cosmology. 
Here we want to show how to define a characteristic unit 
of action for the individual consituents of bound,
stable dynamical systems of global dimension $R$,
made of $N$ elementary constituents. 
By this approach,
it is possible to find that for stable macroscopic systems such an
action is of order $h$ \bib{demartino}. 
Our goal is to show that
any stable gravitationally bound system, where the overall 
relevant interaction 
between constituents is gravity, has exactly
$h$ as characteristic 
unit of action per constituent, if we consider theories of
induced gravity with running Newtonian constant.
What we mean by ``constituents'' in the context of
cosmology are stars or even galaxies. 
We will focus on this application in the next section. 
Let us first illustrate the general procedure, whose
detailed derivation appears in \bib{demartino}.

Let $F(r)$ be the modulus of a classical law of force 
(overall attractive) acting on the $N$
elementary components of mass $m$ which constitute a macroscopic physical 
system. Let $v$ be the some mean local 
characteristic velocity of each individual 
constituent and $\tau$ 
the associated local characteristic time. 
The characteristic mean unit of action\footnote{The characteristic 
action per constituent is conceptually different from the 
interaction action of
field theories used in previous section. For this reason, we use now the
symbol $\alpha$ instead of ${\cal A}$.} per constituent
can be defined as
\beq
\label{9}
\alpha\cong mv^2\tau\,,
\eeq
or, alternatively as
\beq
\label{10}
\alpha\cong rF(r)\tau\,.
\eeq 
The second expression 
is preferable if it is difficult to determine the 
mass and the characteristic velocity of the particle 
(\ie to evaluate the characteristic kinetic energy).
Eq.(\ref{10}) defines an impulsive unit of action 
associated to the time variation of the impulse. 
Eqs.(\ref{9}) and (\ref{10}) do not tell us anything 
about the stability of the system.

A further hypothesis is that if our system is stable,
it obeys the virial theorem in the mean (it is worthwhile to 
stress that any gravitationally bound system, like a galaxy, can be 
dynamically treated only under this hypothesis \bib{binney}). 
Then the mean potential energy of a particle must be of the same order 
of magnitude of its mean
characteristic kinetic energy. In other words,
\beq
\label{11}
{\l}\cong mv^2\,,
\eeq
is the mean characteristic work performed by the system on a
single constituent. 
Since the system is virialized, we can write
\beq
\label{12}
{\l}\cong NF(R)R\,,
\eeq
where $R$ is the mean length scale of the system, of the 
order of magnitude of its global space extension.
The mean characteristic velocity per constituent 
can then be written as
\beq
\label{13}
v\cong\sqrt{NF(R)Rm^{-1}} \, .
\eeq

On the other hand, 
considering the global size $R$ of the system and 
the characteristic global unit of time
of the system ${\cal T}$ (that is the characteristic time for the evolution 
of the system as a whole) we can also write\footnote{For 
a galaxy, ${\cal T}$ is the time 
after which the system evolves and become stable. It is of the order 
of 10 Gyr.}
\beq
\label{14}
v\simeq\frac{R}{{\cal T}}\,.
\eeq
By combining the two previous expressions, 
the characteristic action per constituent can then be written
in the form \bib{demartino}:
\beq
\label{15}
\alpha\cong m^{1/2}R^{3/2}\sqrt{F(R)}\frac{\sqrt{N}}{\cal T}\tau\,.
\eeq
Now, following Calogero \bib{calogero}, 
let us consider $\tau$ 
as the characteristic time associated to the 
local chaotic component of the motion that each constituent 
undergoes due to the force $F(r)$ exherted by all other 
constituents in the system. Such collective chaotic
effect can be modeled, according to Calogero, by 
some time--statistical fluctuation, which can be mathematically
expressed as
\beq
\label{16}
\tau\cong\frac{\cal T}{\sqrt{N}}\,,
\eeq
Consequently, the 
characteristic unit of action 
becomes independent of the number of constituents,
as well as of the global and local characteristic
unit of time \bib{demartino}:
\beq
\label{17}
\alpha\cong m^{1/2}R^{3/2}\sqrt{F(R)}\,.
\eeq
On the other hand, by the fluctuative hypothesis
(\ref{16}), we can write
\beq
\label{18}
\alpha\cong \epsilon\tau\,,
\eeq
where $\epsilon$ is the characteristic energy per particle given by
\beq
\label{19}
\epsilon\cong\frac{E}{N}\,.
\eeq
$E$ is the total energy of the system and 
\beq
\label{20}
A\cong E{\cal T}\,,
\eeq
is the total action. As a consequence, the total action is related to the
characteristic action by the relation
\beq
\label{21}
\alpha\cong N^{-3/2}A\,.
\eeq
Let us now shortly describe an example of computation
of the characteristic unit of action for a stable 
macroscopic system. For a detailed analysis, and a 
comprehensive treatment of several other macroscopic systems
see \bib{demartino}.

Let us consider a stable bunch of charged particles in an a particle
accelerator. Confinement and stability are due to the 
interactions among the
constituents and between the constituents and the external electromagnetic 
fields. The net effect can be schematized 
by saying that the single charged 
particle experiences an effective harmonic force (when higher anharmonic
contributions can be neglected). 
The classical law of force is then 
$F(R)\cong KR$ where $K$ is the effective phenomenological elastic constant.
Then, the characteristic unit of action for each beam constituent is
\beq
\label{23}
\alpha\cong m^{1/2}R^{2}K^{1/2}\,.
\eeq
Inserting typical numbers (\eg transverse oscillations of proton at HERA),
$K\cong 10^{-9}$g sec$^{-2}$, $R\cong 10^{-5}$cm, $m_{p}\cong 10^{-24}$g,
we obtain $\alpha\cong h$ (the same result holds for electrons but 
$K\cong 10^{-8}$g sec$^{-2}$, and we have to take $m_{e}$ instead 
of $m_{p}$). The same results are obtained considering beam data
from the other currently existing accelerators.

The same techniques can be applied
 to several other bound electromagnetic
systems ranging from atoms to molecular clusters, to plasmas 
in quasi--equilibrium, to 
Bose--Einstein condensats and to other systems
at mascroscopic and mesoscopic scales. 
In all cases, the computation of the
characteristic unit of action yields Planck action constant
$h$ \bib{demartino}.
We can say that all these classical and 
semiclassical systems can be described by
classical mechanics plus a suitable classical fluctuation
which mimicks the fundamental quantum structure 
in yielding a characteristic unit of action of order 
$h$. In some sense, $h$ is the quantum signature of the system. 

\section{\normalsize \bf The characteristic unit of 
action for gravitationally bound systems}

The case of gravity is more subtle due to the intrinsic difference of such
an interaction with respect to the others. As we have said, the absence
of screening mechanisms allows it to act practically at all scales. However,
its intrinsic weakness ($\sim 10^{40}$ times weaker than 
electromagnetic forces) makes it efficient in forming
bound structures only if the other interactions can be neglected. On the 
other hand, at early epochs, gravity was comparable or stronger than the
other forces.

As was shown by Calogero \bib{calogero},  
a characteristic action involving
Newtonian interactions can be easily constructed. 
If $m=m_{p}\simeq 10^{-24}$g
is the typical mass of the proton (considering nucleons
to be the granular constituents of the 
Universe), $R\simeq 10^{28}$cm can be assumed as the size 
of the observed Universe, 
and if we assume the Newton law of force for $F(r)$, we get 
\beq
\label{24}
\alpha\cong G_{N}^{1/2}m^{3/2}R^{1/2}\,,
\eeq
which yields $\alpha\simeq h$. From this result, Calogero suggests that the
origin of quantization could be attributed to the interaction of every
particle with the background gravitational force due to all other particles
in the Universe. Such a background interaction generates a chaotic component
in the motion of each single particle, with a characteristic time 
$\tau\simeq 10^{-21}$s measuring the time scale of stochasticity
(zitterbewegung). Essentially, Calogero assumed a weak field limit where the
total gravitational energy goes as $R^{-1}$ and relativistic corrections are 
completely neglected. His point of view is completely 
Machian and gravity is considered as the fundamental interaction. 

Eq.(\ref{24})  was derived
starting from (\ref{18}) where the characteristic time is constructed putting
into (\ref{16}) the age of the Universe and the expected number of baryons
as derived by nucleosynthesis \bib{kolb}. The energy per particle $\epsilon$,
as in Eq.(\ref{19}), is given by dividing the total gravitational energy by 
the number of constituents. It is clear that a sort of underlying
Large Numbers Hypothesis philosophy \bib{barrow} is assumed and calculations 
are carried out without taking into account any specific cosmological model.

Comparing Eq.(\ref{24}) with Eqs. (\ref{20}) and (\ref{21}), we have
\beq
\label{25}
\alpha\cong \left(\frac{M}{N}\right)^{3/2}\left(G_{N}R\right)^{1/2}\,.
\eeq
We will show how to recover this fundamental equation by applying
the scheme of Calogero and De Martino {\it et al.} in the
context of the theories of induced gravity.
We have seen in Sec.2, that quantum field theory asks
for a ``fundamental quantum mechanical nature'' of the Universe \bib{dewitt}
capable of inducing the today observed gravitational and cosmological 
constants (without specific assumptions on  the initial conditions) just 
as a consequence of its dynamical behavior. This argument is supported
also by quantum cosmology \bib{halliwell},\bib{carugno}.

This point of view is radically different from that of Calogero since the
fundamental quantum nature should be recognized at any scale. Microscopic
and macroscopic systems, galaxies, clusters,
and other large scale structures should
show such a quantum signature. 

Microscopic systems naturally exhibit quantum signature since $h$ is present 
in any physical quantity connected with them. The characteristic action for 
an atom is trivially $h$. For mesoscopic and macroscopic scales, quantum
coherent phenomena are  
the evidence that quantum mechanics acts also at this level.  
 
In cosmology, the result by Calogero shows that the whole 
Universe, considered as a gravitationally bound system, shows a quantum 
signature. The questions now are: 
Is this true for any gravitationally bound 
system? Does the gravitational coupling play a role in this picture? Can the
absence of screening effects be connected to the fact that quantum gravity
fluctuations act at all scales?

As starting point, let us consider again the result (\ref{24}) by Calogero.
At astrophysical and cosmological
large scales, the granular constituents of a system can be considered the 
stars or the galaxies and not a simply inchoerent distribution of 
baryons\footnote{We are assuming that the main part of the mass in the 
Universe is clustered in stellar--like aggregates; also the so--called MACHOs
(Massive Astrophysical Compact Halo Objects), recently found by microlensing
technique\bib{nature}, have masses of the order of solar mass.}.
As we have said, stars (and planets) are not properly  
simple gravitationally bound systems
since electromagnetic and 
nuclear interactions contribute (with gravity) to the
stability of the system. Then gravity loses its preminent role to bind
a star contrary to what happens, for example, 
in a galaxy. For this reason, stars can 
be considered the granular units of the universe. A globular 
cluster ($\sim 10^6$ stars), a galaxy ($\sim 10^{11}$ stars),
a cluster of galaxies ($\sim 10^{13}$ stars), or a supercluster of galaxies
($\sim 10^{17}$ stars) are typical systems completely bound by gravity where
the other interactions are negligible, and whose elementary constituents
are stars (However, there is an important 
difference between the globular clusters and the other systems: The former
can be considered collisional systems while the other are collisionless 
systems \bib{binney}. This fact changes completely the dynamical treatment
of the two classes of objects).

We have that a typical Main Sequence star has a mass
\beq
\label{26}
M_{s}\simeq 1 M_{\odot}=1.99\times 10^{33}\mbox{g}\cong 1.19\times 10^{57}
\mbox{protons}\,,
\eeq
and then
\beq
\label{27}
m_{p}=m=\frac{M_{s}}{N}\cong 10^{-24}\mbox{g}\,,
\eeq
which is the mass in Eq.(\ref{24}). Now let us take into account a typical 
galaxy. Its mass is
\beq
\label{28}
M_{g}\simeq 10^{11}M_{\odot}\cong 1.19\times 10^{68}\mbox{protons}\,,
\eeq
so that Eq.(\ref{27}) is again recovered. The situation, as obvious, is 
exactly the same for globular clusters, clusters and superclusters of 
galaxies. Then the ratio 
\beq
\label{29}
\frac{M}{N}\simeq 10^{-24}\mbox{g}\,,
\eeq
is the same for any gravitationally bound system.

Let us now compute the characteristic unit of action 
considering, at first, the gravitational
coupling simply as the Newton constant $G_{N}$. A typical galaxy like
the Milky Way has a linear size of the order $R\simeq 30$kpc (a parsec is
approximatively 1pc=$3.1\times 10^{18}$cm), a mass of the order $M_{g}\simeq
10^{11}M_{\odot}$. As previously seen, the number of baryons is 
$N\simeq 10^{68}$. Introducing these numbers into Eq.(\ref{25}), we get
\beq
\label{30}
\alpha\simeq 10^{-28}\mbox{erg s}\,.
\eeq
Considering now a typical cluster of galaxies with linear size
$R\simeq 10$Mpc and $M_{c}\simeq 10^{13}M_{\odot}$, we obtain
\beq
\label{31}
\alpha\simeq 10^{-27.5}\mbox{ erg s}\,.
\eeq
Finally, at supercluster scales, \ie $R\simeq 100$Mpc and 
$M_{sc}\simeq 10^{17}M_{\odot}$, we obtain
\beq
\label{32}
\alpha\simeq 10^{-27}\mbox{erg s}\simeq h\,,
\eeq
that is approximatively the Planck constant value.
All these values must be taken with an error of an order of magnitude.

It is interesting to observe that the result of Calogero for
the whole Universe is better recovered the larger is
the structure considered.
However, we have to take care of the large 
uncertainties with which cosmological quantities are known. In fact, the 
order of magnitude of the Hubble radius is 
$R\simeq 3\times 10^3\tilde{h}^{-1}$Mpc
with $0.4\leq \tilde{h}\leq 1$ depending on the value of the Hubble constant
$H_{0}$ which, actually, is very controversial \bib{peebles}.

In any case, ranging from galaxies to very large scale structures, as 
superclusters, any gravitationally bound system has 
an associated characteristic unit of action per constituent whose
value is very close to the Planck constant. For very large structures, the 
characteristic action completely coincides with $h$. 

We now wish to show that, assuming a running 
gravitational coupling which decreases
with distance (as supposed by Dirac since cosmological times and distances
are  related), we can show that any gravitationally bound system has a
characteristic action coinciding with $h$.

To obtain such a result, we have to replace Eq.(\ref{25}) with
\beq
\label{33}
\alpha=\left(\frac{M}{N}\right)^{3/2}\left[G(R)R\right]^{1/2}\,,
\eeq
which is in the same line of Eq.(\ref{17}).
The ratio $M/N$ is always of the order $\sim 10^{-24}$g, the other term
must be
\beq
\label{34}
\left[G(R)R\right]^{1/2}
\simeq 10^{9}\mbox{cm}^{2}\mbox{s}^{-1}\mbox{g}^{-1/2}\,.
\eeq
For very large scales we must have that $G(R)\rightarrow G_{N}$ which,
besides, has to coincides with the value measured inside the Solar System and
at laboratory level \bib{gillies}. The product $G(R)R$ is the strength of 
the gravitational interaction which is scale--dependent \bib{goldman}.

Reproducing $h$ (at any scale) could be considered the quantum signature
for gravitationally bound systems at any cosmological scales.
Exploiting the laws of variation obtained by the theories of
induced gravity 
in Sec.2, it is easy to reproduce the constraint (\ref{34}).
For example, by Eq.(\ref{4}), if we choose the set of numbers
$\eta=-11$, $r_{0}=10$kpc and $\chi=1/30$, we recover $\alpha\simeq h$ for a 
typical galaxy of size $R\simeq 30$kpc. It is worthwhile to note that the
parameter $r_{0}\simeq 10$kpc is often used to reproduce the flat rotation
curves of spiral galaxies \bib{goldman},\bib{sanders}.
It could be recovered also  taking into consideration the emittance which
is a scale of length (or of ``temperature'') associated to a correlated
system \bib{demartino}. Such a quantity is defined as
\beq
{\cal E}\simeq \lambda_{c}\sqrt{N}\,,
\eeq
where $\lambda_{c}=h/mc$ is the Compton length.
It is connected to the characteristic action \bib{demartino}.
If we consider the Compton length of the proton and the number $N$ of protons
in a galaxy, we obtain ${\cal E}\simeq 10$kpc, that is the emittance
is connected to the typical scale length of the galaxy. In other worlds,
the quantum parameter $\lambda_{c}$ and the number of microscopic
constituents $N$
determine the astrophysical size ${\cal E}$.

Similar results, can be achieved  also using exponential laws as (\ref{7}) or 
(\ref{8}). However, the parameters (and, in some sense, the right law) depend
on the scale which we are considering and only observations can fix exactly
the model. 

The situation is very close to that of ordinary quantum systems: Given a set 
of quantum numbers, we obtain a stable state. In this case, given a set of
gravitational--quantum numbers, we obtain a gravitationally bound stable
system. This could be the simple explaination why gravitationally bound 
systems do not form at any scale.

As a further remark, we can say that also the problem of dark matter is not
so dramatic if one agrees with this picture. The large amount of gravitating
material which people observe at any scale is only due to the fact that
Newton law is always used without considering the change of the strength of
the gravitational interaction.

A last point of controversy is if gravity increases or decreases with the 
scale (see for example \bib{bertolami},\bib{accetta},\bib{deglinnocenti}).
Actually, the renormalized one--loop quantum gravity models forecast both the
options depending on the parameters of the renormalization group equations
\bib{gellmann},\bib{tonin}. Only astrophysical observations and fine
laboratory experiments on the variation of $G(R)$ will decide what is the 
real situation \bib{gillies}.
 
\section{\normalsize \bf Conclusions}
In this paper, we argued that any large scale bound system, where gravity is
the overall interaction among the components, has a characteristic unit
of action which coincides, in order of magnitude, with the Planck constant.
The result is achieved by asking for the variation of gravitational coupling
with scale. This issue is in agreement with quantum gravity models and
with the fact that a screening mechanism is absent in gravity. 

Also if a similar result for the whole Universe was achieved by Calogero,
his point of view is different since, in his picture, gravity is the
fundamental interaction that, in a Machian way, produced a sort of
stochastic quantization. In our case, gravity is induced by quantum field 
theory and the Universe ``at all scales'' has a fundamental quantum 
mechanical nature. However, both approaches ask for a quantum signature
which can be achieved at large scales and, also if we are considering
a variable gravitational coupling, the effects of such a variation are small
(They are confined into one or two order of magnitude. Besides, the
expected variation of Newton constant into the solar system is estimated
to be $\dot{G}/G\sim 10^{-11}$ years$^{-1}$ which is very small 
\bib{gillies}).

If this approach is fundamentally sensible, 
questions like the formation of galaxies, the
dark matter problem, and the role of gravity as a 
fundamental interaction have to be deeply reconsidered.

\vspace{2. mm}

{\bf Aknowledgments}\\
\noindent The authors would like 
to thank Francesco Guerra for very useful
discussions and comments which allowed to 
greatly improve the content of the present paper.

One of us (F. I.) aknowledges the Alexander von
Humboldt--Stiftung for financial support, and the 
Quantum Theory Group of Prof. Dr. Martin Wilkens
at the Fakult\"at f\"ur Physik der Universit\"at Potsdam
for hospitality while on leave of absence from the
Dipartimento di Fisica dell'Universit\`a di Salerno.

\vspace{2. cm}

\begin{centerline}
{\bf REFERENCES}
\end{centerline}
\begin{enumerate}
\item\label{kolb}
E. W. Kolb and M. S. Turner, {\it The Early Universe}
(Addison--Wesley, New York, 1990).
\item\label{sakharov}
A. Sakharov, Zh. Eksp. Teor. Fiz. {\bf 49}, 245 (1965).\\
V. F. Mukhanov, H. A. Feldman and R. H. Brandenberger, 
Phys. Rep. {\bf 215}, 203 (1992).
\item\label{hucra}
V. de Lapparent, M. J. Geller and J. P. Huchra, 
Astrophys. Jour. {\bf 302}, L1 (1986).\\
M. J. Geller and J. P. Hucra, Science {\bf 246}, 897 (1989).
\item\label{calogero}
F. Calogero, Phys. Lett. {\bf A 228}, 335 (1997).
\item\label{demartino}
S. De Martino, S. De Siena and F. Illuminati,
Mod. Phys. Lett. {\bf B 12}, 291 (1998).\\
S. De Martino, S. De Siena and F. Illuminati, {\it Inference 
of Planck action constant by a classical fluctuative
postulate holding for stable microscopic and macroscopic
dynamical systems}, E--preprint quant-phxxx (1998), 
submitted to J. Phys. {\bf A}.\\
N. Cufaro Petroni, S. De Martino, S. De Siena and F. Illuminati, 
{\it A stochastic model for the semiclassical collective
dynamics of charged beams in particle accelerators},
E--preprint physics/9803036 (1998), to appear in the
Proceedings of the International Workshop on
``Quantum Aspects of Beam Dynamics'', held in Stanford,
4--9 January 1998.
\item\label{dirac}
P. A. M. Dirac, Nature {\bf 139}, 323 (1937).\\
P. A. M. Dirac, Proc. Royal Soc. {\bf 165}, 199 (1938).
\item\label{barrow}
J. D. Barrow and F. J. Tipler, {\it The antropic 
cosmological principle} (Oxford University Press, 
Oxford, 1986).
\item\label{cimento}
S. Capozziello, R. de Ritis, C. Rubano and P. Scudellaro,
Rivista del Nuovo Cim. {\bf 4}, 1 (1996).
\item\label{berman}
M. S. Berman, Int. J. Theor. Phys. {\bf 31}, 1217 (1992).
\item\label{beesham}
A. Beesham, Int. J. Theor. Phys. {\bf 33}, 1935 (1994).
\item\label{sciama}
D. W. Sciama, Mon. Not. Royal Ast. Soc. {\bf 113}, 34 (1953).
\item\label{jordan}
P. Jordan, {\it Schwerkraft und Weltall} (Braunschweig: Vieweg, 1955).\\
P. Jordan, Z. Phys. {\bf 157}, 112 (1959).
\item\label{brans}
C. Brans and R. H. Dicke, Phys. Rev. {\bf 124}, 925 (1961). 
\item\label{smolin}
L. Smolin, Nucl. Phys. {\bf B 160}, 253 (1979).\\
A. Zee, Phys. Rev. Lett. {\bf 42}, 417 (1979).\\
S. Adler, Phys. Rev. Lett. {\bf 44}, 1567 (1980). 
\item\label{cqg}
S. Capozziello and R. de Ritis, Class. Quantum Grav. {\bf 11},
107 (1994).
\item\label{stelle}
K. S. Stelle, Phys. Rev. {\bf D 16}, 953 (1977).\\
K. S. Stelle, Gen. Rel. Grav. {\bf 9}, 353 (1978).
\item\label{fradkin}
E. S. Fradkin and A. A. Tseytlin, Nucl. Phys. {\bf B 201}, 469 (1982).
\item\label{avramidy}
E. G. Avramidy and A. O. Barvinky, Phys. Lett. {\bf B 159}, 269 (1985).
\item\label{birrell}
N. D. Birrell and P. C. W. Davies, {\it Quantum Fields in Curved Space}
(Cambridge University Press, Cambridge, 1982).
\item\label{amendola}
L. Amendola, S. Capozziello, M. Litterio and F. Occhionero, 
Phys. Rev. {\bf D 45}, 417 (1992).
\item\label{la}
D. La P. J. Steinhardt, Phys. Rev. Lett. {\bf 62}, 376 (1989).
\item\label{capozziello}
S. Capozziello and R. de Ritis, Phys. Lett. {\bf A 177}, 1 (1993). \\
S. Capozziello, R. de Ritis and C. Rubano Phys. Lett. {\bf A
177}, 8 (1993).
\item\label{weinberg}
S. Weinberg, Phys. Rev. {\bf 140}, B516 (1965).\\
B. De Witt, Phys. Rev. {\bf 162}, 1239 (1967).
\item\label{goldman}
T. Goldman, J. P\'erez--Mercader, F. Cooper and M. M. Nieto, 
Phys. Rev. {\bf B 281}, 219 (1992).
\item\label{gellmann}
M. Gell--Mann and F. Low, Phys. Rev. {\bf 95}, 1300 (1954).
\item\label{tonin}
J. Julve and M. Tonin, Nuovo Cimento {\bf B 46}, 137 (1978).
\item\label{bertolami}
O. Bertolami, J. M. Mourao and J. P\'erez--Mercader, 
Phys. Rev. {\bf B 311}, 27 (1993).
\item\label{sanders}
R. H. Sanders, Astron. Astrophys. Rev. {\bf 2}, 1 (1990).
\item\label{eckhardt}
D. H. Eckhardt, Phys. Rev. {\bf D 48}, 3762 (1993).
\item\label{3kpc}
M. Capaccioli, S. Capozziello, R. de Ritis and G. Longo,
Astron. Nachr. {\bf 318}, 1 (1997).
\item\label{binney}
J. Binney and S. Tremaine, {\it Galactic Dynamics}
(Princeton University Press, Princeton, 1987).
\item\label{dewitt}
J. A. Wheeler, Rev. Mod. Phys. {\bf 29}, 463 (1957).\\
B.S. de Witt, Phys. Rev. {\bf 160}, 1113 (1967).
\item\label{halliwell}
J. J. Halliwell in {\it Quantum Cosmology and Baby Universes} ed. S. Coleman 
et al. (World Scientific, Singapore, 1991).
\item\label{carugno}
E. Carugno, S. Capozziello and F. Occhionero, Phys. Rev. {\bf D 47}, 
4261 (1993).
\item\label{nature}
E. Aubourg {\it et al.}, Nature {\bf 365}, 623 (1993).\\
C. Alcock {\it et al.}, Nature {\bf 365}, 621 (1993).
\item\label{peebles}
P. J. E. Peebles, {\it Principles of Physical Cosmology}
(Princeton University Press, Princeton, 1993).
\item\label{gillies}
A. J. Sanders and G. T. Gillies,
Rivista del Nuovo Cim. {\bf 2}, 1 (1996).
\item\label{accetta}
F. S. Accetta, L. M. Krauss, and P. Romanelli, Phys. Rev. 
{\bf B 248}, 146 (1990).
\item\label{deglinnocenti}
S. Degl'Innocenti, G. Fiorentini, G. G. Raffelt, B. Ricci and A. Weiss,
Astron. Astrophys. {\bf 312}, 345 (1996). 
\end{enumerate}
 
\vfill 
\end{document}